\pdfoutput=1

\documentclass[11pt]{article}

\usepackage[final]{acl}

\usepackage{times}
\usepackage{latexsym}
\usepackage{subcaption}
\usepackage{textcomp}
\usepackage{fmtcount}
\usepackage{svg}
\usepackage{graphicx}  
\usepackage{booktabs}
\usepackage{xcolor}

\usepackage[T1]{fontenc}

\usepackage[utf8]{inputenc}

\usepackage{microtype}

\usepackage{inconsolata}

\usepackage{graphicx}
\usepackage{booktabs}
\title{Typographic Attacks in a Multi-Image Setting}

\author{Xiaomeng Wang \\
  Radboud University \\
  \texttt{xiaomeng.wang@ru.nl} \\\And
  Zhengyu Zhao \\
  Xi'an Jiaotong University\\
  \texttt{zhengyu.zhao@xjtu.edu.cn} \\\And
  Martha Larson \\
  Radboud University \\
  \texttt{m.larson@cs.ru.nl}}

\begin{document}
\maketitle
\begin{abstract}
Large Vision-Language Models (LVLMs) are susceptible to typographic attacks, which are misclassifications caused by an attack text that is added to an image.
In this paper, we introduce a multi-image setting for studying typographic attacks, broadening the current emphasis of the literature on attacking individual images.
Specifically, our focus is on attacking image sets without repeating the attack query.
Such non-repeating attacks are stealthier, as they are more likely to evade a gatekeeper than attacks that repeat the same attack text.
We introduce two attack strategies for the multi-image setting, leveraging the difficulty of the target image, the strength of the attack text, and text-image similarity.
Our text-image similarity approach improves attack success rates by 21\% over random, non-specific methods on the CLIP model using ImageNet while maintaining stealth in a multi-image scenario. 
An additional experiment demonstrates transferability, i.e., text-image similarity calculated using CLIP transfers when attacking InstructBLIP.

\end{abstract}

\section{Introduction}

\begin{figure}[t]
    \centering
    \begin{subfigure}[t]{0.48\columnwidth}
        \centering
        \includegraphics[width=\linewidth]{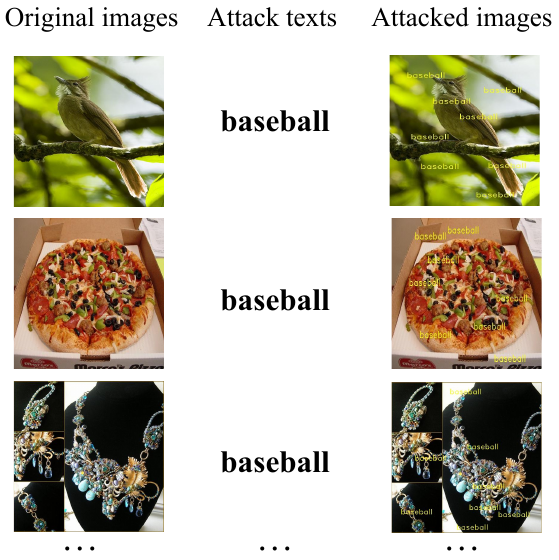} 
        \caption{Repeating}
        \label{fig:other_strategy_multi-image_setting}
    \end{subfigure}
     \hfill
    \begin{subfigure}[t]{0.48\columnwidth}
        \centering
        \includegraphics[width=\linewidth]{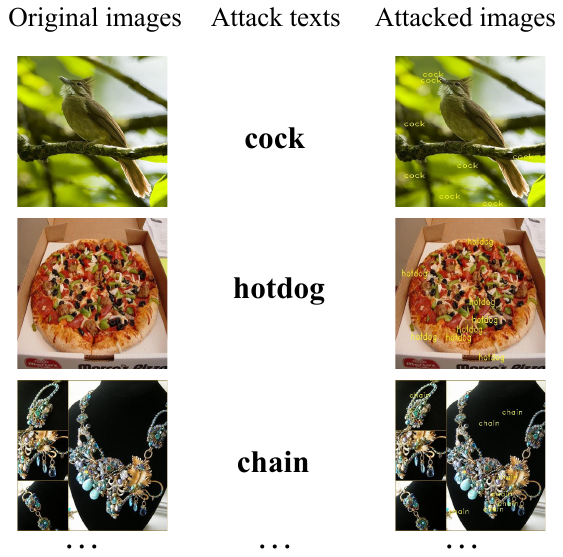} 
        \caption{Non-repeating (ours)}
        \label{fig:our_strategy_multi-image_setting}
    \end{subfigure}
        \caption{In real-world attack scenarios, an attacker would target a CLIP-based system with a set of images rather than a single image.
        The use of a repeating text (as in a) more strongly signals a typographic attack than the use of diversified texts (as in b).
        In this paper, we introduce the multi-image attack setting, which limits attack text repetition, and we show the importance of text-image similarity when choosing an attack text for a given target image.
        }
        \vspace{-2mm}
    \label{fig:multi-image_setting}
\end{figure}

Large Vision-Language Models (LVLMs), such as the contrastive language-image pretraining model (CLIP)~\citep{clip} and InstructBLIP~\citep{dai2023instructblip}, have shown remarkable performance across a variety of multimodal downstream tasks.
However, a small, but important, set of research contributions has recently demonstrated that LVLMs are vulnerable to typographic attack on their classification abilities~\citep{multi_neurons,typo_mllms,self_typo}.
In typical typographic attacks, a text consisting of one or more words is added to an image, superimposed in the middle~\citep{typo_mllms} or added at the top or bottom~\citep{self_typo}. Alternatively, it is added as a physical label to an object before the target image is taken~\citep{multi_neurons}.
A typographic attack is successful when the text, which we refer to as the \emph{attack text}, leads to the misclassification of the target image.
Typographic attacks exploit the capability of LVLMs to interpret not only the visual content of images but also any written language that they contain.
In this paper, we introduce the multi-image attack setting for studying typographic attacks and the importance of the similarity between the attack text and target image.

We explain the multi-image attack setting with the help of Figure~\ref{fig:multi-image_setting}.
In previous works~\citep{typo_mllms,multi_neurons}, the emphasis is on attacking individual images, and the same attack text could be used repeatedly for different target images (Figure~\ref{fig:other_strategy_multi-image_setting}).
In our work, in contrast, we consider the pattern of attack texts used for the attack across a set of images.
Specifically, we are interested in the case in which the attack does not repeat across the images (Figure~\ref{fig:our_strategy_multi-image_setting}).
In short, studying an attack in a multi-image attack setting means studying the attack from a holistic perspective.
The importance of non-repeating attacks is elaborated in Section~\ref{sec:evading}. 

Developing typographic attacks for the non-repeating multi-image setting requires answering a key question: What strategy should the attacker use to attack the set of target images in a way that maintains the attack success rate but avoids repeating attack texts?
This paper reveals that the attacker should take the similarity between the attack text and the target image into account. 

First, we look at a possible alternative to text-image similarity, namely, \emph{attack text effectiveness} (ATE).
Researchers have found that the targeted attack success rate varies with the attack text~\citep{multi_neurons}.
ATE is the average attack success of an attack text measured across a set of images.

Next, we study how images behave under attack.
The literature has reported that adversarial attacks are not uniformly effective across all images~\citep{image_selection}.
We build on this finding and study \emph{visual image prediction probability} (VIPP), i.e., how challenging it is to predict the class of an image.
Using VIPP, we can prioritize images and propose attack strategies in which the most difficult images are assigned attack texts first.

Then, we turn to the strategies proposed for typographic attacks in the multi-image setting.
Our first type of strategy tests the contribution of using ATE to select attack texts for images prioritized by VIPP. 
Our second type of strategy tests the advantages of using text-image similarity to select attack texts, with and without VIPP image-prioritization.

Our paper makes the following contributions:
\vspace{-2mm}
\begin{itemize}
\vspace{-2mm}
    \setlength\itemsep{-0.2em}
    \item We introduce and explain the importance of the multi-image setting for typographic attacks and of studying non-repeating typographic attacks within this setting. 
    \item We identify and discuss the importance of text-image similarity for typographic attacks. 
    \item We propose two types of strategies for typographic attacks and test them in our non-repeating multi-image setting.
    \item We carry out an analysis that provides insights into the trade-off between the number of times attack texts are repeated in the multi-image setting and the attack success rate. 
    \item We demonstrate that text-image similarity calculated with respect to CLIP will generalize when attacking another model (InstructBLIP).
\end{itemize}

Source code for this paper is available. \footnote{\url{https://github.com/XiaomengWang-AI/Typographic-Attacks-in-a-Multi-Image-Setting}}

\section{Background and Motivation}
In this section, we provide further details on the background of and motivation for our work.

\subsection{Importance of the multi-image setting}

We are interested in studying the multi-image setting because it gives us insight into how real-world systems might be attacked.
The issue is becoming increasingly important as CLIP and other LVLMs incorporating its pretrained vision encoder are being used as the basis for more applications, where misclassifications have serious real-world consequences.
For example, CLIP has been used in systems for the detection of unknown objects on roads~\citep{ Bogdoll22RoadObjects}, hateful content~\citep{ Gonzalez-Pizarro23hateful}, fake news~\citep{ Tahmasebi23FakeNews} and illegal outdoor advertising~\citep{ Zhang24OutAds}.
In this paper, we do not develop attacks on classifiers with serious real-world contexts, rather we use ImageNet data and the associated classes to study attacks.

\subsection{The nature of the misclassification threat}
When a target image is attacked with an attack text, it is pushed into the embedding space in the direction of the semantics of that attack text.
The push can be quite strong such that the image is misclassified into the class with the same target label as the attack text.
Until now, the literature has been mainly focused on this case and reported the attack success rate (ASR) for so-called \emph{targeted} misclassifications~\citep{multi_neurons,typo_mllms,self_typo}. 
However, in our work, we observe that in a non-negligible number of cases, the label of the class into which the image is misclassified is not identical to the attack text.  
Since in critical real-world scenarios, any misclassification is potentially harmful (e.g., one roadsign being recognized as another) we focus on measuring ASR over all misclassifications, which we refer to as the \emph{untargeted} ASR.

\subsection{Stealth: Evading the gatekeeper}
\label{sec:evading}
We envision that such applications would use a \emph{gatekeeper} (human or machine) to monitor the incoming image input (i.e., the set of images submitted to the system). 
Within the multi-image setting, non-repeating attack texts (Figure~\ref{fig:our_strategy_multi-image_setting}) are important to study because the lack of repetition improves the \emph{stealth} of the attack, i.e., the incoming image input appears less suspicious to the gatekeeper. 

The gatekeeper cannot block all incoming images that contain text because text in images is important in application domains.
In fact, in the widely-used LAION2B dataset, 50\% of the images have been reported to contain text~\citep{Lin24Parrot}. 
However, the gatekeeper may become suspicious when the same text is used over again on different images.
A similar observation has been made in the literature on perturbative adversarial images.
Adding Universal Adversarial Perturbations (UAP)~\citep{moosavi2017universal,sandovalUnlearnable} can cause a misclassification. However, once the UAP pattern is known, the inspector can easily reverse the adversarial images~\citep{sandoval2022poisons,sandoval2022autoregressive}.


\subsection{Motivation for text-image similarity}
A typographic attack pushes an image away from its original position in the embedding space to a position that no longer correctly reflects the semantics of the image’s visual content.
We anticipate that the easiest attack is one that achieves semantic impact with only a small change in the image position, i.e., an attack that moves an image into a neighboring semantic class. 
Where the training of the LVLM is ideal, semantically similar classes will lie close to each other in the model’s embedding space.
Where the training of the LVLM is not ideal, classes that the model can easily confuse will lie close to each other.
In both cases, the attack texts with the highest similarity to the target image are those attack texts corresponding to neighboring classes.
For this reason, we expect attack texts with high text-image similarity to be the most effective attack texts.
Certainly, the embedding space should not be envisioned as a set of mutually exclusive semantic classes, but the same reasoning holds if we understand the embedding space to be characterized by semantic regions or regions of confusion.

\section{Related Work}
\subsection{Contrastive Language-Image Pretraining}
Contrastive Language-Image Pretraining (CLIP)~\citep{clip} is a large vision-language model trained from scratch using a contrastive learning objective on a dataset comprising 400 million image-text pairs collected from the internet. 
CLIP is designed to learn representations of images alongside their corresponding paired texts to align these representations from the two modalities within the same embedding space.
This alignment ensures that corresponding image-text pairs are closer in the embedding space compared to non-corresponding pairs.
After training, the vision encoder of CLIP learns to associate images with their corresponding paired texts. 
This capability enables CLIP to excel at zero-shot transfer tasks across various domains, such as image classification, optical character recognition, and semantic segmentation. 

\subsection{Typographic attacks against LVLMs}
Recent studies~\citep{multi_neurons, azuma2023defense, noever2021reading} show that typographic attacks can impair the zero-shot classification capability of CLIP. 
~\citet{multi_neurons} claims that the underlying reason for typographic attacks could be multimodal neurons responding to shared concepts across different formats.
Other LVLMs, such as InstructBLIP~\citep{dai2023instructblip} and LLaVA~\citep{llava}, are expected to inherit similar typographic weaknesses when incorporating the vision encoder of CLIP. 
Studies in~\citet{self_typo, typo_mllms} evaluate the robustness of LVLMs to typographic attacks, including InstructBLIP, LLaVA 1.5, MiniGPT4-2, and GPT4-V models.
~\citet{typo_mllms} selects the attack text by random method, further evaluating the impact of font size, color, opacity, and spatial positioning on the typographic attack success.
~\citet{self_typo} proposes novel typographic attacks, termed Self-Generated Attacks, which leverage the capabilities of LVLMs to identify visually similar deceiving classes or generate descriptive reasoning for more effective deception.
This work~\citep{self_typo} is close to our work because the prompt to the LVLMs requests an attack text that is `similar' to the target image.
In contrast, our work calculates text-image similarity directly in the embedding space. 
As a result, our attacks are more directly related to confusion regions arising due to shortcomings in the training of LVLMs, laying the groundwork for future study of principled defenses.
As already mentioned, our work differs from previous contributions in our focus on the multi-image setting.


\section{Insights on Typographic Attacks}
\label{sec:insights}

In this section, we carry out a set of typographic attacks that allows us to analyze attack texts, target images, and text-image similarity.
The analysis reveals the underlying factors that have an impact on the success rate of typographic attacks and provides insights, on which we based our proposed attack strategies for the multi-image setting (cf. Section~\ref{sec:attack}).

\subsection{Target model, data, and attack settings}
\label{sec:evaluation_setting}
In our analysis in this section and our experiments (Section~\ref{sec:experiments}), we study OpenCLIP~\citep{open_clip} as the target model, which is an open-source implementation of CLIP.
We attack the zero-shot classification task.
We use the publicly available OpenCLIP (ViT-B/32) trained on the LAION2B dataset~\citep{laion}.
We use a server with the following configuration: a 20GB memory NVIDIA A10 GPU.

Our \emph{evaluation dataset} is built from the ILSVRC2012 validation dataset~\citep{imagenet}.
It consists exclusively of images that have been correctly classified by the target model, which is a total of 30,940 of the original 50,000 images.
Evaluating with only correctly classified images is conventional in research on adversarial images~\citep{TabacofExploring,narodytska2017simple}, because it ensures that the attack success rate (ASR) includes only misclassifications that are direct results of attacks.

Two design choices in our analysis and experiments are related to the fact that we are studying the patterns of typographic attacks in the multi-image setting and not the form of the attack.
First, we focus on attack texts that consist of only one word, leaving the study of multi-word attacks to future work.
The set of attack texts used in this paper is drawn from the class labels of the ILSVRC2012 validation dataset.
It consists of 579 unique words, corresponding to the 579 labels from the original 1000 labels that consist of only one word.

Second, we adopt the font style recommendations for attack texts from the previous work~\citep{typo_mllms}. 
The font color is set as yellow, and the opacity is 100\%. 
The font size is adjusted to 0.8 times the font type size. 
Regarding the spatial positioning of the attack text, we follow the protocol in~\citet{multi_neurons} to use the same eight arbitrarily chosen coordinates and maintain a consistent font style.
As seen in Figure~\ref{fig:multi-image_setting}, the attack texts are readily evident to human inspection.
Recall that, in this paper, our focus is stealth as related to the non-repetition of attack texts and not the inconspicuousness of attack texts.

We report untargeted ASR and targeted ASR, as previously mentioned.
An attack is successful if the top-1 prediction label of the attacked image is different from the ground-truth label of the original image.
For targeted ASR, only the cases in which the top-1 prediction label is identical to the attack text label are calculated.
Targeted ASR reflects the extent to which the attack text is able to pull the target images into its exact semantic direction.
Note that the level of attack success rate that an attack needs to achieve in order to be considered dangerous depends on the domain. 
In some domains, even a few images evading the gatekeeper could cause a serious problem.
In this work, we compare attack success rates without interpreting them in a particular real-world scenario.

\subsection{Attack text effectiveness} 
\label{sec:ATE}
\begin{figure}
    \centering
    \includegraphics[width=\columnwidth]{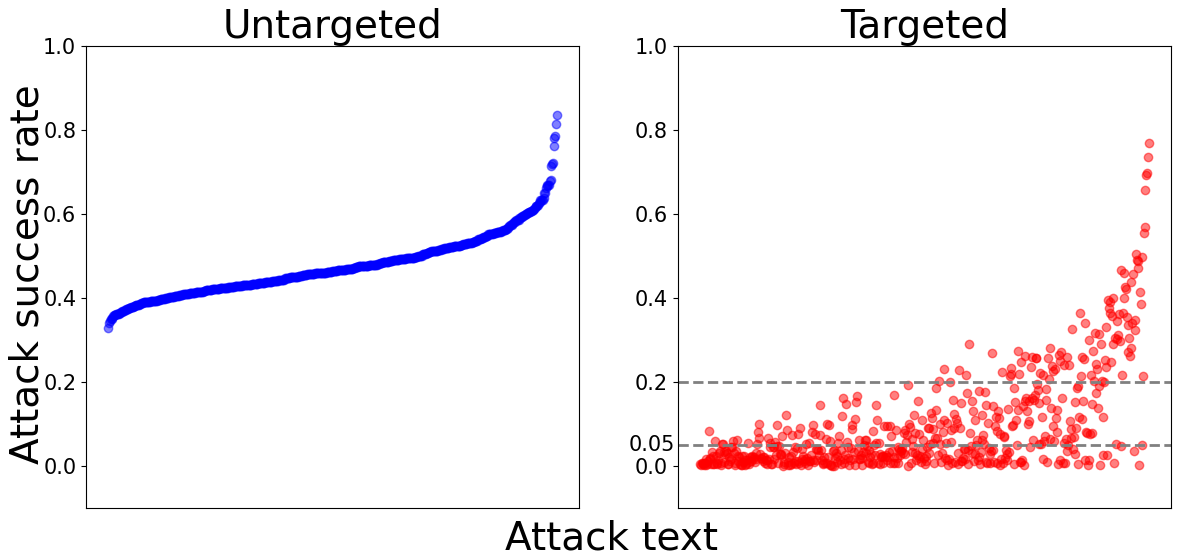}
    \caption{Attack success rates of our 579 attack texts. The attack texts are arranged in ascending order of untargeted attack success rate along the horizontal axis.}
    \label{fig:attack_text_esults}
\end{figure}

Not all texts are equally suited for carrying out typographic attacks.
We define \emph{attack text effectiveness} as the property of a text that reflects this suitability. 
Following the methodology in~\citet{multi_neurons}, we conduct a brute-force attack on all images in the evaluation dataset using the same attack text.
For each attack text, we calculate the attack success rates based on these attacks. 
Specifically, higher attack success rates indicate better attack text effectiveness.
Figure~\ref{fig:attack_text_esults} shows the attack success rate for each attack text.
We can observe that for both with respect to untargeted and targeted ASRs, there is a large amount of variation between texts in terms of their ability to cause misclassification.

\subsection{Prioritizing images} 
\label{sec:VIPP}
In this section, we study a property of images that we refer to as the visual image prediction probability (VIPP).
It is defined as the prediction probability of the ground-truth label for the original image. 
A higher prediction probability of the ground-truth label indicates that the model is more likely to identify this image correctly.

\begin{figure}
    \centering
    \includegraphics[width=\columnwidth]{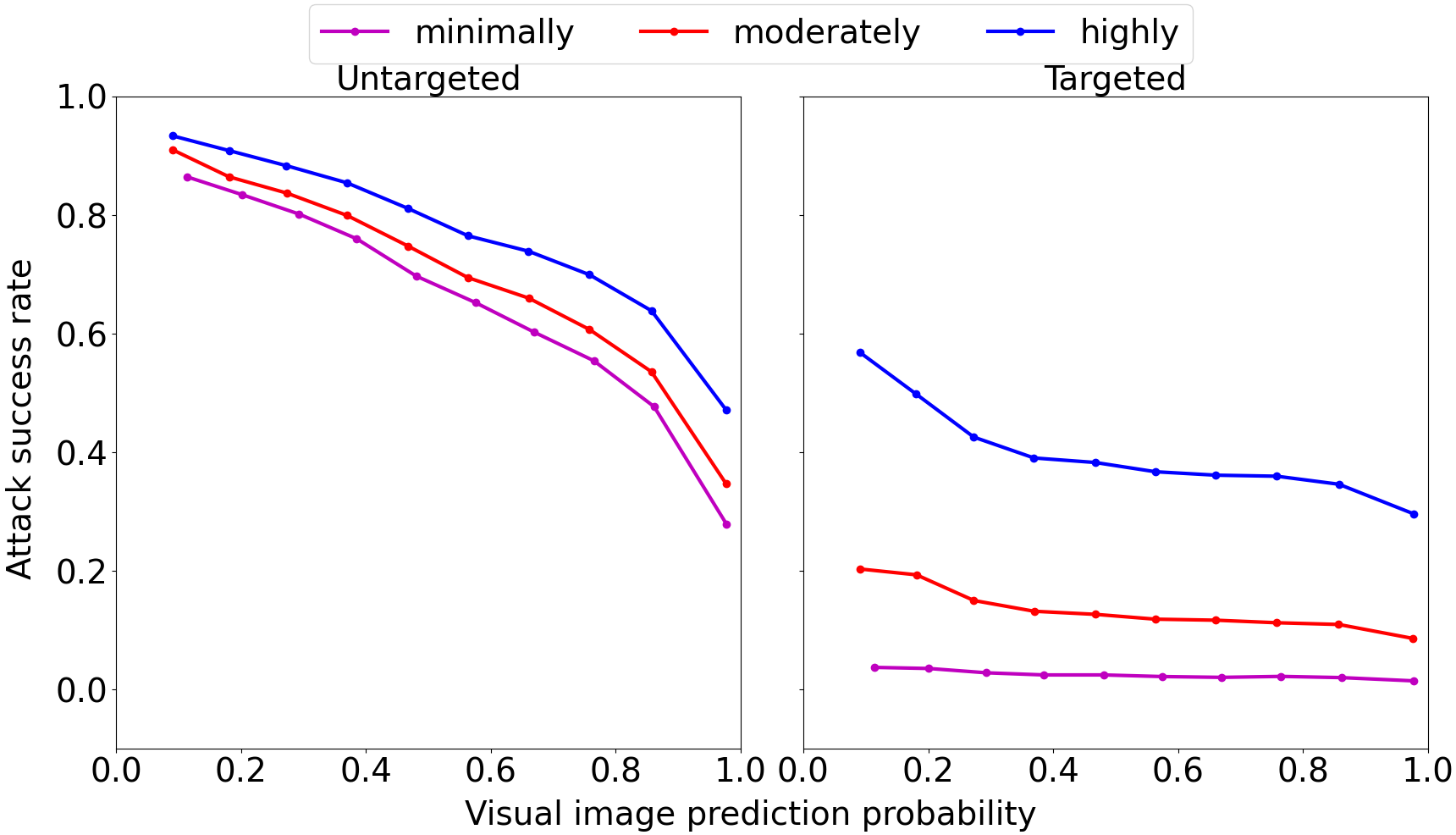}
   \caption{Attack success rates versus visual image prediction probability for three categories of attack texts: highly, moderately, and minimally effective. }
    \label{fig:vid_asr}
\end{figure}

We are interested in the relationship between VIPP and the attack success rates.
To get a detailed understanding of this relationship, we analyze this relationship for three separate types of attack text: highly, moderately, and minimally effective.
Figure~\ref{fig:vid_asr} shows the trend of the attack success rates as the VIPP gets higher.
We see that indeed difficult-to-classify images (high VIPP) are more difficult to attack (low ASR).
This observation holds across all three types of attack text.
However, the performance of minimally effective attack texts measured with respect to targeted ASR is particularly low.
This graph suggests that it is useful to prioritize images using VIPP and use the most effective attack texts to attack the most difficult images.

Here, we provide some details on how Figure~\ref{fig:vid_asr} was produced.
The three categories of attack texts were created on the basis of ASR by choosing thresholds.
The grey lines in the targeted ASR plot of Figure~\ref{fig:attack_text_esults} show the thresholds that we chose (minimally effective is lower than 0.05 ASR, moderately effective is between 0.05 and 0.2 ASR, and highly effective is larger than 0.2).
Note that we are interested in general trends and did not optimize these categories (i.e., they could have been chosen based on different criteria or based on untargeted ASR).
We divided the images into equally spaced bins based on their VIPP and excluded bins containing less than 35 images.
For each of the categories of attack text, we use each attack text it contains to attack all images in each bin.
For each bin, we then have one attack text ASR for each attack text, which we average to a bin-level ASR and plot.

\subsection{Text-image interactions} 
\label{sec:text-image}

The text-image similarity is a measure that quantifies the degree of interaction between the attack text and the target image embedding.
We use the cosine similarity between the embeddings of the target image and the attack text, both encoded by the CLIP model.
Note that since the training of CLIP is not necessarily ideal throughout the embedding space, the text-image similarity does not necessarily correspond to semantic similarity. 

\begin{figure}
    \centering
    \includegraphics[width=\columnwidth]{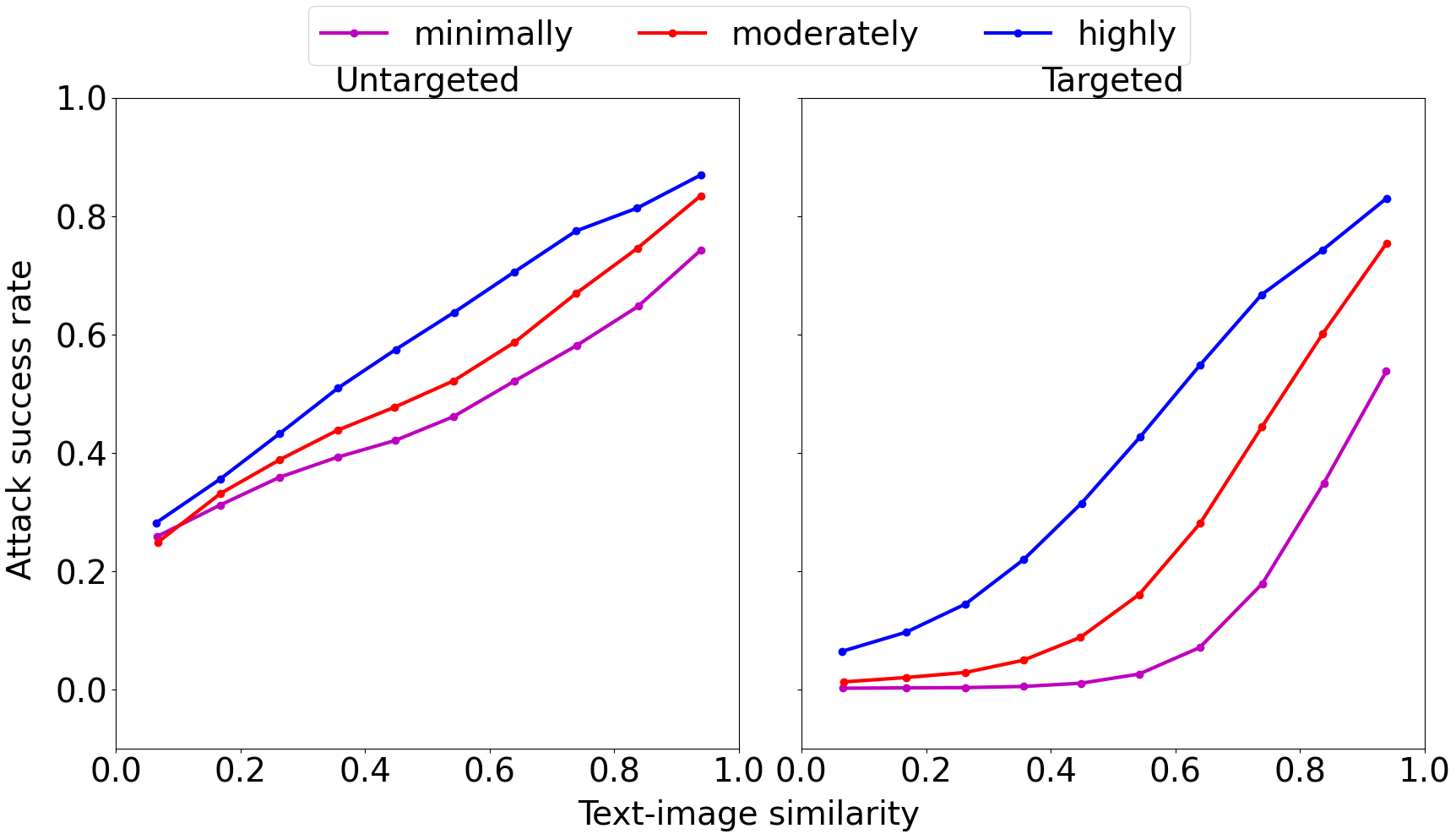}
    \caption{Attack success rates versus text-image similarity for three categories of attack texts: highly, moderately, and minimally effective. 
    The attack success rates generally increase as text-image similarity rises.}
    \label{fig:total_sim_asr}
\end{figure}

Figure~\ref{fig:total_sim_asr} shows the relationship between text-image similarity and attack success rates for the three types of attack text, minimally, moderately, and highly effective.
For all three types of attack text, attacks become more successful as the similarity grows higher.
This graph suggests that it is useful to use text-image similarity to choose which attack text to use to attack a given image.
Figure~\ref{fig:total_sim_asr} is generated in the following way.
We first attack all target images using each attack text in each category while calculating the text-image similarity.
The similarities are then equally grouped into 10 bins, and the results for each bin are averaged.
In the next section, we will introduce the two types of attack strategies that we propose based on our analyses in this section.

\section{Attack Strategies}
\label{sec:attack}

In this section, we first provide the specifics of how we instantiate the multi-image attack setting with non-repeating attack texts for our experiments. 
Then, we describe the attack strategies we propose on the basis of the insights in Section~\ref{sec:insights}.

\subsection{Non-repeating multi-image attack}
To carry out our experiments, we conceptualize a non-repeating multi-image attack, represented by Figure~\ref{fig:set_attack_diagram} as a matching problem. 
The attacker has a set of target images to attack and a set of attack texts that can be used to attack.
For each target image, the attacker must choose an attack text, without repeating.
The goal of the attack is to cause a misclassification, such that the target image is no longer classified as its ground-truth label.
In our experiments, the set of images and the set of attack texts have the same size, making the attack a 1-to-1 matching problem. We compare our proposed strategies against a baseline approach.

\noindent\textbf{Rand: Random.} In this baseline approach, one target image and one attack text are randomly selected from their respective sets. 
Each attack text may be selected only once.

\subsection{Attack text effectiveness strategies}
We propose two attacks that make use of attack text effectiveness (ATE), presented in Section~\ref{sec:ATE}.
We conjecture that the best matching strategy is one for which the most difficult-to-attack images (reflected by high VIPP) are attacked with the strongest attack texts (reflected by high ATE).
To gain insight into the importance of matching difficult-to-attack target images with strong attack images, we propose a strategy that reverses the relationship to measure the impact of the attack.

\noindent\textbf{HighVIPP-LowATE: VIPP and Attack Text Effectiveness.} Lower Success Rate Strategy: In this approach, target images are sorted by descending visual image prediction probability (VIPP), and attack texts are sorted by ascending targeted attack success rate. 
Each target image is then matched one-to-one with an attack text in sequence, aligning images with higher prediction probabilities with texts that have lower attack success rates.

\noindent\textbf{HighVIPP-HighATE: VIPP and Attack Text Effectiveness.} Higher Success Rate Strategy: For this strategy, target images are again sorted by descending VIPP, but attack texts are sorted by descending targeted attack success rate. 
Consequently, each target image is matched one-to-one with an attack text in sequence, this time, the images with higher prediction probabilities are matched with the text with higher attack text effectiveness.

The ATE-based strategies provide an alternate way of matching attack texts and target images to compare with the Text-image similarity strategies, which we describe next.

\subsection{Text-image similarity strategies}
We propose two attacks that make use of text-image similarity, presented in Section~\ref{sec:text-image}.

\noindent\textbf{Rand-TextImSim: Text-image similarity.} In this approach, each target image is assigned an attack text based on the highest text-image similarity (excluding the ground-truth class label). 
All images are ordered randomly, and for each, the text-image similarity is calculated to determine the best text match.
The drawback of this approach is that if the closest attack text to a given target image has previously been used for another image it is no longer available. 
There is no mechanism that can save the attack texts for the images where they are most needed.
To address this, we propose also using VIPP prioritization with text-image similarity.

\begin{figure}
    \centering
    \includegraphics[width=\columnwidth]{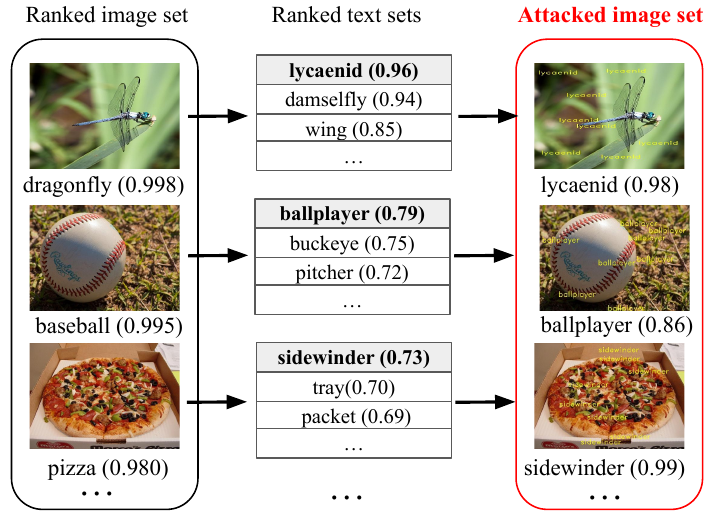}
    \caption{One-to-one matching between the image and attack text sets. The image and text sets are ranked by descending VIPP and text-image similarity. The attack text is selected from the ranked list (excluding the ground truth label and previously used texts). The text and numeric values under each image represent its prediction label and probability, while the bracketed values for each attack text indicate text-image similarity.}

    \label{fig:set_attack_diagram}
\end{figure}

\noindent\textbf{VIPP-TextImSim: VIPP and text-image similarity.} 
The approach is illustrated in Figure~\ref{fig:set_attack_diagram} and consists of the following steps.
    
\noindent {\it 1.} We sequence the target images by visual image prediction probability (VIPP) in descending order to prioritize images on which it is most challenging to conduct an attack.

\noindent {\it 2.} Starting with the top image in the prioritized list, we calculate the text-image similarity between the image and all of our attack texts.

\noindent {\it 3.} Moving down the list, for each target image, we select the attack text with the highest text-image similarity after the image's own ground-truth label and attack text labels that have already been used have been excluded. 

In the next section, we report the results of our experiments that apply these attack strategies.

\section{Experiments}
\label{sec:experiments}
\subsection{Experimental setting}
\label{sec: experimental setting}
The model and attack settings for the experiments are the same as were used in the analysis in Section~\ref{sec:insights} and were previously described in Section~\ref{sec:evaluation_setting}.
In this section, we continue to use the same 579 one-word attack texts.
We test our model on five randomly drawn sets of target images containing 579 images each.
Recall that the evaluation data contains only images that are correctly classified by our model, OpenCLIP.
To draw one test set of 579 images we follow the following procedure.
We chose the 579 test images by first randomly choosing 579 classes from the original 1000 classes of the LSVRC2012 validation dataset~\citep{imagenet}.
Then, we randomly choose one image from each of the 579 classes.
This method provides us with a satisfying degree of certainty that the target images are semantically well-balanced. 
Note that the outcome of our experiments would be less informative if either the 579 attack texts or the 579 target images were highly homogenous with respect to their semantics.

\subsection{Non-repeating attacks in the multi-image setting}
\label{sec:non-reapting attacks}
\begin{figure}
    \centering
    \includegraphics[width=\columnwidth]{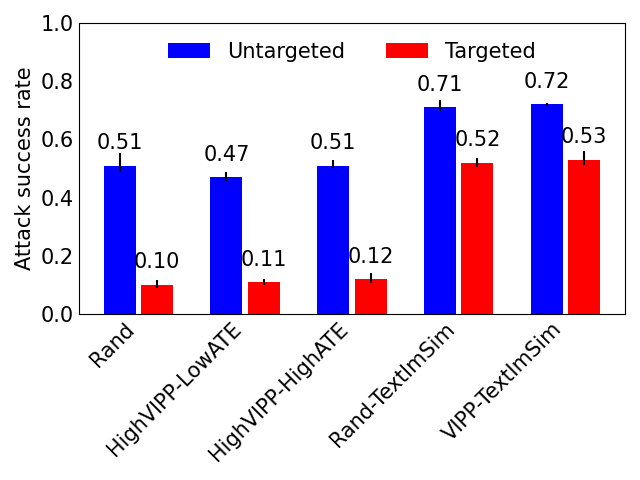}
    \caption{Comparisons of the five strategies in the multi-image setting. 
    The values above the bars denote the average attack success rates.
    The error bars represent the variability of the attack success rates.
    VIPP-TextImSim strategy achieves the highest attack success rates.}
    \label{fig:set_based_attack}
\end{figure}

The evaluation results, depicted in Figure~\ref{fig:set_based_attack}, reveal the importance of text-image similarity, with Rand-TextImSim and VIPP-TextImSim being the highest performers by a substantial margin.
The two approaches, however, achieve comparable ASR, i.e., VIPP prioritization is not making a measurable contribution.
Further, HighVIPP-HighATE performs closely to Rand, our random baseline.

We conclude that using ATE for attack text selection is not competitive with using TextImSim in the multi-image attack setting.
Recall, however, that the experimental setup requires one-to-one matching, making it quite challenging.
For this reason, we should not abandon VIPP or ATE entirely.
Evidence that these properties are potentially impactful is the difference between the two ATE approaches.
Specifically, when we reverse the use of ATE, matching difficult images (HighVIPP) with weak attack texts (LowATE), we observe a deterioration of performance over HighVIPP-HighATE.

It is interesting to observe that for the text-image similarity strategies, a greater proportion of the untargeted ASR is comprised of targeted ASR, reflecting a large number of misclassifications being made into the class whose label is identical to the attack text.
Referring to Figure~\ref{fig:set_attack_diagram} allows us to gain further insight into how text-image similarity selects attack texts for images.
These examples are real examples of cases in which the attack causes a targeted misclassification.
In the top example, the relationship between the target image and the attack text is one of visual similarity, between the target image of the dragonfly and the attack text \emph{lycaenid}, which is a type of butterfly, both of which are winged insects.
In the middle example, the relationship is one of semantic similarity, with the baseball in the target image being a key object associated with the attack text \emph{ballplayer}.
At the bottom, the relationship is again visual, but corresponds to a glitch in the semantic space, with the pizza in the target image being associated with the attack text \emph{sidewinder}, which is a kind of snake.
Future work should focus on understanding how typographic attacks take advantage of regions of the CLIP embedding space in which the representation of semantics is not optimal.
Also, future work should investigate the untargeted cases, in which the image is misclassified into a class unrelated to the attack text.
These cases have the potential to be particularly stealthy because the intent of the attacker cannot be read from the text in the image.

\section{Loosening the Non-repeating Requirement}
\begin{figure}
    \centering
    \includegraphics[width=\columnwidth]{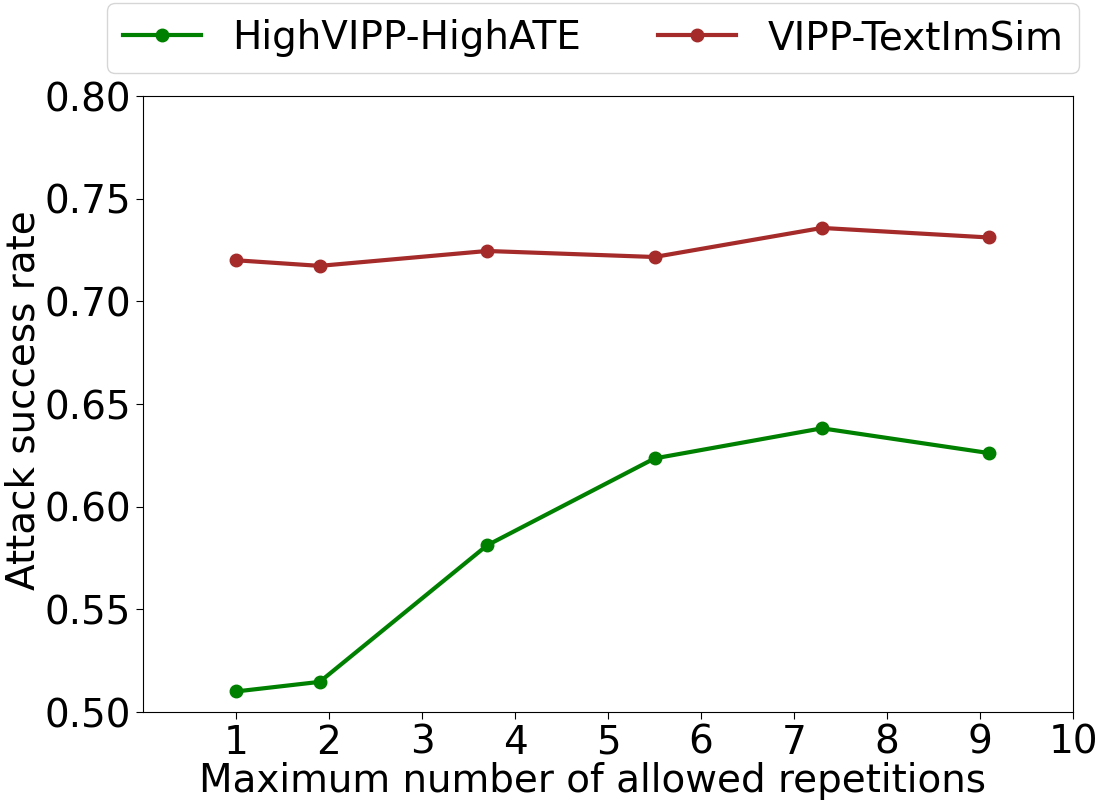}
    \caption{Untargeted attack success rates versus the maximum number that a given attack text is allowed to repeat. VIPP-TextImsim achieves the highest untargeted attack success rate as the number of allowed repetitions grows from 1 to 10.}
    \label{fig:method35_repetition_untargeted_asr}
\end{figure}

In this section, we gradually increase the maximum number of repetitions of attack texts allowed in the multi-image setting.
We aim to study the trade-off between the stealth of the multi-image attack and the ASR.
We focus on HighVIPP-HighATE and VIPP-TextImSim, which represent our ATE and TextImSim approaches.
Figure~\ref{fig:method35_repetition_untargeted_asr} plots the untargeted ASR as we allow attack texts to repeat an increasing number of times from the non-repeating setting (which corresponds to 1 at the far left).
Targeted ASR is not depicted, but follows the same trends.

The plot reveals that the strength of the VIPP-TextImSim strategy with respect to the HighVIPP-HighATE strategy is maintained as more and more repetitions are allowed, supporting the importance of text-image similarity for typographic attacks.
The ASR of both VIPP-TextImSim and HighVIPP-HighATE increase as more repetitions are admitted, reflecting the trade-off between stealth (i.e., restricted attack text repetitions) and attack strength (i.e., ASR).
However, HighVIPP-HighATE increases more steeply, which we attribute to the increased use of attack texts with the highest effectiveness, i.e., those on the far right of Figure~\ref{fig:attack_text_esults}.

We note that the performance of VIPP-TextImSim is limited by the attack texts that are available in the set of 579 attack texts used in this experiment.
The set might fail to obtain optimal attack texts, meaning attack texts with the highest possible similarity to the target image.
Future work should study the impact of the set of candidates from which the attack text can be chosen on approaches using text-image similarity.

\section{Evaluation on Other LVLMs}
\label{sec:evaluation on LVLMs}

In the previous sections, we have experimented with a whitebox attack on CLIP. 
We assume that attackers have full access to the trained CLIP model that they are attacking.
This experimental setting is not unrealistic because many LVLMs are publicly available.
However, it is interesting to understand whether our multi-image attack can also be carried out without whitebox access.
In this section, we perform a greybox attack on the InstructBLIP (FlanT5XXL)~\citep{dai2023instructblip} model using VIPP-TextImSim, our best-performing method from Section~\ref{sec:experiments}.
The attacker does not access InstructBLIP to create the attack but instead has access to CLIP.
We refer to this attack as greybox and not blackbox, since InstructBLIP uses a vision
encoder similar in structure to CLIP and its training dataset has a substantial overlap with the CLIP training set.

We choose the InstructBLIP (FlanT5XXL) 
because it demonstrates better optical character recognition (OCR) performance, e.g., compared to the MiniGPT4 model evaluated in~\citet{ocrbench}.
The OCR performances mean that InstructBLIP focuses on text in images and is for this reason particularly susceptible to typographic attack~\citep{self_typo}.
Future work can also look at the LLaVA~\citep{llava} model, which exhibits limited robustness against typographic attacks~\citep{typo_mllms}, as well as models without strong OCR performance.

\subsection{Experimental setting}
\begin{figure}[t]
    \centering
    \begin{subfigure}[t]{\columnwidth}
        \centering
        \includegraphics[width=\linewidth]{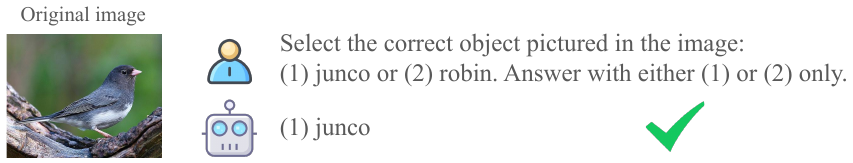} 
        \caption{Model output for the original image.}
        \label{fig:original case}
    \end{subfigure}
     \vfill
    \begin{subfigure}[t]{\columnwidth}
        \centering
        \includegraphics[width=\linewidth]{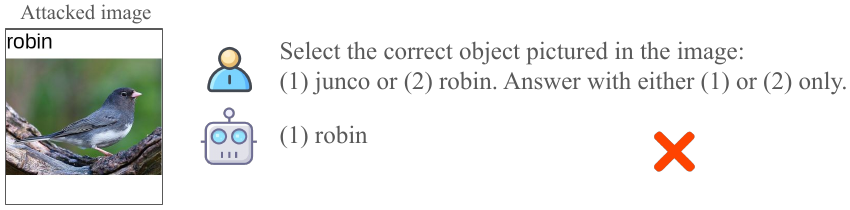} 
        \caption{Model output for the attacked image.}
        \label{fig:attack case}
    \end{subfigure}
        \caption{Model outputs for the original and attacked images given the multiple options in the prompt.
        }
    \label{fig:original and attack case}
\end{figure}

Our experiments follow the setting introduced in Section~\ref{sec: experimental setting} but integrate modifications that are necessary for testing attacks on LVLMs.
We first make the choice of a prompt to use, cf. Figure~\ref{fig:original and attack case}.
To choose the prompt, we follow~\citet{self_typo}, a recent study on single-image typographic attacks, which includes an attack on InstructBLIP.
The prompt used in~\citet{self_typo} is `Select the correct \{Dataset Subject\} pictured in the image: (1) \{first option\} or (2) \{second option\}. Answer with either (1) or (2) only.'
Like~\citet{typo_mllms}, this prompt gives the model a small, closed set of options (only two).
In order to evaluate untargeted ASR, it is necessary for the model to output one of the full set of options (1000 ImageNet class labels).
During our exploratory experiments, we found that InstructBLIP cannot deal well with a longer list.
Also, posing an open question adds the step of mapping the output back to the full set of options, which can introduce error. 
For this reason, we decided to stick closely to~\citet{self_typo}, adding only a sentence to explicitly request one answer only. 

Our final prompt is, `Select the correct object pictured in the image: (1) \{ground-truth label\} or (2) \{attack text label\}. Answer with either (1) or (2) only.'
We shuffle the ordering of each option to avoid model bias to any answer order.
In future work, more focus on prompt engineering, addressing the issue of the high sensitivity of models to prompts, would allow the evaluation of untargeted attacks. 
However, in this section, we will evaluate only the targeted case.

As in~\citet{self_typo}, the attack text is placed on the top white space in black font color and size of 20px, as shown in Figure~\ref{fig:attack case}.



\subsection{LVLM attack in the multi-image setting}

A comparison of the Rand and our VIPP-TextImSim attacks is shown in Table~\ref{tab:evaluation results}.
The `Original Image Set' column shows the original accuracy of InstructBLIP.
Since either attack strategy specifies its own \{attack text label\}, the two attacks may have different original accuracy, although this difference turns out to have little impact (0.62 vs. 0.63).
We also see the original accuracy is relatively low, consistent with that observed in~\citet{self_typo} (for other datasets).
The `Attacked Image Set' column shows the performance of the attacks.
Our VIPP-TextImSim method achieves a significant drop in accuracy (from 0.63 to 0.39) on the attacked image set, compared to the random baseline (from 0.62 to 0.49). 

Additionally, we point to the difference in the style of the text added to the image for the typographic attack.
When attacking the CLIP model, text was added eight times scattered over the image, as in Figures~\ref{fig:multi-image_setting} and~\ref{fig:set_attack_diagram}.
When attacking InstructBLIP, the text is added more consistently with how text occurs naturally in a social image collection (at the top of an image), as in Figure~\ref{fig:attack case}.
In the future, more work on natural-looking attack texts can further increase the stealth of typographic attacks, both in the single-image and multi-image settings. 

\begin{table}[]
    \centering
    \resizebox{0.5\textwidth}{!}{ 
   \begin{tabular}{ccc}
\toprule
Attack         & Original Image Set & Attacked  Image Set \\ \midrule
Rand           & 0.62             & 0.49              \\
VIPP-TextImSim & 0.63             & 0.39              \\ 
\bottomrule
\end{tabular}
    }
    \caption{Accuracy of InstructBLIP with respect to the original ground-truth label on the image set used in Section.~\ref{sec:non-reapting attacks} (original and attacked images). Note: 0.49 and 0.39 correspond to targeted ASRs of 0.51 and 0.61.}
    \label{tab:evaluation results}
\end{table}

\section{Conclusion and Outlook}
In this paper, we have introduced a multi-image attack setting for typographic attacks and studied the contribution of the similarity between target images and attack text to typographic attacks.
We have argued that repeated use of the same attack text is not the most dangerous attack, since it would be evident to a gatekeeper.
Once stealth is taken into account, the importance of text-image similarity for typographic attacks becomes clear.

Moving forward, we foresee that typographic attacks based on text-image similarity can reveal issues in the embedding space of LVLMs.
Also, research is needed to understand the factors influencing attack texts that cause untargeted misclassifications, investigating whether the distraction introduced by the text is semantic or purely visual in nature.
Such investigations can improve LVLMs, making them robust to the dangers of typographic attacks. 
Studying diverse attack texts in the multi-image setting will continue to be part of this effort.

\section*{Limitations}
Our work is primarily limited in two dimensions.
First, we focus on single-word attack texts, excluding the potential complexities introduced by multi-word texts.
In real-world scenarios involving more complex and varied visual texts with images, the findings of this paper may differ.
Future work could expand the types of texts used in attacks to assess whether the observed trend holds for more sophisticated attack texts.
Second, compared with imperceptible perturbative attacks, the way that we superimposed the attack text on the image when attacking CLIP is more conspicuous, thereby limiting attack stealth at the single image level.
Future work may explore more natural methods of placing the attack text, while ensuring semantic consistency with the image content, similar to images in the real world, thus reducing suspicion at the image level.

\section*{Ethical Considerations}
This paper was inspired by concerns about typographic attacks on real-world detection systems based on CLIP, such as those used for detecting hateful content. 
We point out the reality of a multi-image setting and the significance of stealth for typographic attacks.
We propose new strategies to conduct typographic attacks on models, such as CLIP and InstructBLIP, more effectively in the multi-image setting.
This technique could potentially be exploited by malicious users.
However, we believe that our work highlights the potential risk of typographic attacks in a multi-image setting, thus inspiring the development of effective defense strategies.
Moreover, in this paper, we do not design or optimize attacks on classifiers used in critical applications.
Instead, we conduct all experiments using ImageNet data and select classes labeled within this dataset as our attack texts.

\section*{Acknowledgement}
This work was carried out on the Dutch national e-infrastructure with the support of SURF Cooperative.
This work was partially supported by the European Union under the Horizon Europe project AI-CODE (GA No. 101135437).

\bibliography{stealthy_typo_attack}

\begin{thebibliography}{24}
\providecommand{\natexlab}[1]{#1}

\bibitem[{Azuma and Matsui(2023)}]{azuma2023defense}
Hiroki Azuma and Yusuke Matsui. 2023.
\newblock Defense-prefix for preventing typographic attacks on clip.
\newblock In \emph{Proceedings of the IEEE International Conference on Computer Vision, Workshop on Adversarial Robustness in the Real World}, pages 3644--3653.

\bibitem[{Bogdoll et~al.(2022)Bogdoll, Eisen, Nitsche, Scheib, and Zöllner}]{Bogdoll22RoadObjects}
Daniel Bogdoll, Enrico Eisen, Maximilian Nitsche, Christin Scheib, and J.~Marius Zöllner. 2022.
\newblock Multimodal detection of unknown objects on roads for autonomous driving.
\newblock In \emph{Proceedings of the IEEE International Conference on Systems, Man, and Cybernetics}.

\bibitem[{Cheng et~al.(2024)Cheng, Xiao, and Xu}]{typo_mllms}
Hao Cheng, Erjia Xiao, and Renjing Xu. 2024.
\newblock Unveiling typographic deceptions: Insights of the typographic vulnerability in large vision-language models.
\newblock In \emph{Proceedings of the European Conference on Computer Vision}.

\bibitem[{Cherti et~al.(2023)Cherti, Beaumont, Wightman, Wortsman, Ilharco, Gordon, Schuhmann, Schmidt, and Jitsev}]{open_clip}
Mehdi Cherti, Romain Beaumont, Ross Wightman, Mitchell Wortsman, Gabriel Ilharco, Cade Gordon, Christoph Schuhmann, Ludwig Schmidt, and Jenia Jitsev. 2023.
\newblock Reproducible scaling laws for contrastive language-image learning.
\newblock In \emph{Proceedings of the IEEE Conference on Computer Vision and Pattern Recognition}.

\bibitem[{Dai et~al.(2023)Dai, Li, Li, Tiong, Zhao, Wang, Li, Fung, and Hoi}]{dai2023instructblip}
Wenliang Dai, Junnan Li, Dongxu Li, Anthony Tiong, Junqi Zhao, Weisheng Wang, Boyang Li, Pascale Fung, and Steven Hoi. 2023.
\newblock Instruct{BLIP}: Towards general-purpose vision-language models with instruction tuning.
\newblock In \emph{Advances in Neural Information Processing Systems}, volume~36, page 49250–49267.

\bibitem[{Goh et~al.(2021)Goh, †, †, Carter, Petrov, Schubert, Radford, and Olah}]{multi_neurons}
Gabriel Goh, Nick~Cammarata †, Chelsea~Voss †, Shan Carter, Michael Petrov, Ludwig Schubert, Alec Radford, and Chris Olah. 2021.
\newblock \href {https://doi.org/10.23915/distill.00030} {Multimodal neurons in artificial neural networks}.
\newblock \emph{Distill}.
\newblock Https://distill.pub/2021/multimodal-neurons.

\bibitem[{González-Pizarro and Zannettou(2023)}]{Gonzalez-Pizarro23hateful}
Felipe González-Pizarro and Savvas Zannettou. 2023.
\newblock Understanding and detecting hateful content using contrastive learning.
\newblock In \emph{Proceedings of the International AAAI Conference on Web and Social Media}.

\bibitem[{Lin et~al.(2024)Lin, He, Wang, Wang, Li, and Shou}]{Lin24Parrot}
Yiqi Lin, Conghui He, Alex~Jinpeng Wang, Bin Wang, Weijia Li, and Mike~Zheng Shou. 2024.
\newblock Parrot captions teach {CLIP} to spot text.
\newblock In \emph{Proceedings of the European Conference on Computer Vision}.

\bibitem[{Liu et~al.(2024{\natexlab{a}})Liu, Li, Wu, and Lee}]{llava}
Haotian Liu, Chunyuan Li, Qingyang Wu, and Yong~Jae Lee. 2024{\natexlab{a}}.
\newblock Visual instruction tuning.
\newblock In \emph{Advances in Neural Information Processing Systems}, volume~36, page 34892–34916.

\bibitem[{Liu et~al.(2024{\natexlab{b}})Liu, Li, Huang, Yang, Yu, Li, Yin, Liu, Jin, and Bai}]{ocrbench}
Yuliang Liu, Zhang Li, Mingxin Huang, Biao Yang, Wenwen Yu, Chunyuan Li, Xu-Cheng Yin, Cheng-Lin Liu, Lianwen Jin, and Xiang Bai. 2024{\natexlab{b}}.
\newblock Ocrbench: on the hidden mystery of ocr in large multimodal models.
\newblock \emph{Science China Information Sciences}, 67(12).

\bibitem[{Moosavi-Dezfooli et~al.()Moosavi-Dezfooli, Fawzi, Fawzi, and Frossard}]{moosavi2017universal}
Seyed-Mohsen Moosavi-Dezfooli, Alhussein Fawzi, Omar Fawzi, and Pascal Frossard.
\newblock Universal adversarial perturbations.
\newblock In \emph{Proceedings of the IEEE conference on Computer Vision and Pattern Recognition}.

\bibitem[{Narodytska and Kasiviswanathan(2017)}]{narodytska2017simple}
Nina Narodytska and Shiva~Prasad Kasiviswanathan. 2017.
\newblock Simple black-box adversarial attacks on deep neural networks.
\newblock In \emph{Proceedings of the Conference on Computer Vision and Pattern Recognition Workshops}.

\bibitem[{Noever and Noever(2021)}]{noever2021reading}
David~A Noever and Samantha E~Miller Noever. 2021.
\newblock Reading isn't believing: Adversarial attacks on multi-modal neurons.
\newblock \emph{arXiv preprint arXiv:2103.10480}.

\bibitem[{Ozbulak et~al.(2021)Ozbulak, Anzaku, De~Neve, and Van~Messem}]{image_selection}
Utku Ozbulak, Esla~Timothy Anzaku, Wesley De~Neve, and Arnout Van~Messem. 2021.
\newblock Selection of source images heavily influences the effectiveness of adversarial attacks.
\newblock In \emph{Proceedings of the British Machine Vision Conference}.

\bibitem[{Qraitem et~al.(2024)Qraitem, Tasnim, Saenko, and Plummer}]{self_typo}
Maan Qraitem, Nazia Tasnim, Kate Saenko, and Bryan~A Plummer. 2024.
\newblock Vision-llms can fool themselves with self-generated typographic attacks.
\newblock In \emph{Advances in Neural Information Processing Systems, Workshop on Multimodal Algorithmic Reasoning}.

\bibitem[{Radford et~al.(2021)Radford, Kim, Hallacy, Ramesh, Goh, Agarwal, Sastry, Askell, Mishkin, Clark et~al.}]{clip}
Alec Radford, Jong~Wook Kim, Chris Hallacy, Aditya Ramesh, Gabriel Goh, Sandhini Agarwal, Girish Sastry, Amanda Askell, Pamela Mishkin, Jack Clark, et~al. 2021.
\newblock Learning transferable visual models from natural language supervision.
\newblock In \emph{Proceedings of the International Conference on Machine Learning}.

\bibitem[{Russakovsky et~al.(2015)Russakovsky, Deng, Su, Krause, Satheesh, Ma, Huang, Karpathy, Khosla, Bernstein, Berg, and Fei-Fei}]{imagenet}
Olga Russakovsky, Jia Deng, Hao Su, Jonathan Krause, Sanjeev Satheesh, Sean Ma, Zhiheng Huang, Andrej Karpathy, Aditya Khosla, Michael Bernstein, Alexander~C. Berg, and Li~Fei-Fei. 2015.
\newblock {ImageNet Large Scale Visual Recognition Challenge}.
\newblock \emph{International Journal of Computer Vision}, 115(3):211--252.

\bibitem[{Sandoval-Segura et~al.(2022{\natexlab{a}})Sandoval-Segura, Singla, Fowl, Geiping, Goldblum, Jacobs, and Goldstein}]{sandoval2022poisons}
Pedro Sandoval-Segura, Vasu Singla, Liam Fowl, Jonas Geiping, Micah Goldblum, David Jacobs, and Tom Goldstein. 2022{\natexlab{a}}.
\newblock Poisons that are learned faster are more effective.
\newblock In \emph{Proceedings of the IEEE Conference on Computer Vision and Pattern Recognition, Workshop on Art of Robustness}.

\bibitem[{Sandoval-Segura et~al.(2023)Sandoval-Segura, Singla, Geiping, Goldblum, and Goldstein}]{sandovalUnlearnable}
Pedro Sandoval-Segura, Vasu Singla, Jonas Geiping, Micah Goldblum, and Tom Goldstein. 2023.
\newblock What can we learn from unlearnable datasets?
\newblock In \emph{Advances in Neural Information Processing Systems}, volume~36, page 75372–75391.

\bibitem[{Sandoval-Segura et~al.(2022{\natexlab{b}})Sandoval-Segura, Singla, Geiping, Goldblum, Goldstein, and Jacobs}]{sandoval2022autoregressive}
Pedro Sandoval-Segura, Vasu Singla, Jonas Geiping, Micah Goldblum, Tom Goldstein, and David Jacobs. 2022{\natexlab{b}}.
\newblock Autoregressive perturbations for data poisoning.
\newblock In \emph{Advances in Neural Information Processing Systems}, volume~35, page 27374–27386.

\bibitem[{Schuhmann et~al.(2022)Schuhmann, Beaumont, Vencu, Gordon, Wightman, Cherti, Coombes, Katta, Mullis, Wortsman et~al.}]{laion}
Christoph Schuhmann, Romain Beaumont, Richard Vencu, Cade Gordon, Ross Wightman, Mehdi Cherti, Theo Coombes, Aarush Katta, Clayton Mullis, Mitchell Wortsman, et~al. 2022.
\newblock {LAION-5B}: An open large-scale dataset for training next generation image-text models.
\newblock In \emph{Advances in Neural Information Processing Systems}, volume~35, pages 25278--25294.

\bibitem[{Tabacof and Valle(2016)}]{TabacofExploring}
Pedro Tabacof and Eduardo Valle. 2016.
\newblock Exploring the space of adversarial images.
\newblock In \emph{Proceedings of the International Joint Conference on Neural Networks}.

\bibitem[{Tahmasebi et~al.(2023)Tahmasebi, Hakimov, Ewerth, and M\"{u}ller-Budack}]{Tahmasebi23FakeNews}
Sahar Tahmasebi, Sherzod Hakimov, Ralph Ewerth, and Eric M\"{u}ller-Budack. 2023.
\newblock Improving generalization for multimodal fake news detection.
\newblock In \emph{Proceedings of the ACM International Conference on Multimedia Retrieval}.

\bibitem[{Zhang et~al.(2024)Zhang, Ding, Sharid Kayes~Dipu, Lv, Huang, Suleiman~Abdullahi, Zhang, Song, and Wang}]{Zhang24OutAds}
Haiyan Zhang, Zheng Ding, Md~Sharid Kayes~Dipu, Pinrong Lv, Yuxue Huang, Hauwa Suleiman~Abdullahi, Ao~Zhang, Zhaoyu Song, and Yuanyuan Wang. 2024.
\newblock Identification of illegal outdoor advertisements based on {CLIP} fine-tuning and {OCR} technology.
\newblock \emph{IEEE Access}, 12:92976--92987.

\end{thebibliography}

\end{document}